\documentclass[prl,twocolumn,showpacs,superscriptaddress]{revtex4}

\usepackage{graphicx}
\usepackage{mathrsfs}
\usepackage{dcolumn}
\usepackage{bm}
\usepackage{amsmath}
\usepackage{amsfonts}
\usepackage{color}

\begin{document}

\title{Confinement Effect Driven Quantum Spin Hall Effect in Monolayer AuTe$_2$Cl}

\author{Guyue Zhong}
\email{gyzhong@hust.edu.cn}
\affiliation{Wuhan National High Magnetic Field Center and School of Physics, Huazhong University of
Science and Technology, Wuhan 430074, China}
\author{Q. Xie}
\affiliation{Wuhan National High Magnetic Field Center and School of Physics, Huazhong University of
Science and Technology, Wuhan 430074, China}
\author{Gang Xu}
\email{gangxu@hust.edu.cn}
\affiliation{Wuhan National High Magnetic Field Center and School of Physics, Huazhong University of
Science and Technology, Wuhan 430074, China}

\date{\today}

\begin{abstract}

Based on first-principles calculations,
we predict that
the monolayer AuTe$_2$Cl
is a quantum spin Hall (QSH) insulator with
a topological band gap about 10~meV.
The three-dimensional (3D)
AuTe$_2$Cl is a topological semimetal
that can be viewed as the monolayer
stacking along ${\bf b}$ axis.
By studying the energy level distribution of $p$ orbitals of Te atoms for
the bulk and the monolayer,
we find that
the confinement effect driven
{$p_y^{-}-p_z^{+}$ }
band inversion
is responsible for
the topological nontrivial nature of monolayer.
Since 3D bulk AuTe$_2$Cl has already been experimentally
synthesized, we expect that
monolayer AuTe$_2$Cl can be exfoliated from a bulk sample
and the predicted QSH effect can be observed.

\end{abstract}

\maketitle
\section{Introduction}
Two-dimensional (2D) topological insulators (TIs),
known as quantum spin Hall (QSH) insulators
are characterized by a topological nontrivial bulk gap
and gapless helical edge states  protected by
time-reversal symmetry from backscattering
at the sample boundaries
~\cite{RevModPhys.82.3045,RevModPhys.83.1057,Yan_2012}.
Because of their potential applications in semiconductor spintronics,
they have recently attracted great attention in condensed matter physics~\cite{ML_2012}.
For example, the QSH edge channels may be
useful for applications in integrated circuit technology,
where power dissipation is becoming a more and more
serious problem as devices becoming smaller.
Up to now, the QSH effect only has been observed
in HgTe/CdTe~\cite{Ber_2006,MK_2013},
InAs/GaSb~\cite{Cl_2008,KI_2011} quantum wells and
WTe$_2$~\cite{WS_2018}
under low temperature and ultra-high vacuum,
which hinder their applications in realistic devices.
To overcome the thermal disturbance and promote
the practical application of QSH insulators,
those materials containing heavy atoms with
extremely strong spin-orbital coupling(SOC),
such as MX (M = Zr, Hf and X = Cl, Br, I)
~\cite{ZL_2015}, MTe$_5$ (M = Zr, Hf)~\cite{WHM_2014},
BiTeI~\cite{MSB_2012},
have been extensively studied.
These results demonstrate that those 2D
materials are good candidates to realize the QSH effect at room temperature.
In spite of those progresses,
desirable materials preferably with high feasibility
of experimental realization are still extremely scarce and deserve to explore in experiment and theory.

Recently, the catalogue of topological electronic materials
greatly enriches the number of new topological
materials, which has listed all 3D TIs and
topological semimetals (TSMs) based
on symmetry indicators~\cite{ZTT_2019,MGV_2019,FT_2019},
and inspires the enthusiasm for the study of new topological materials.
As we know, some 3D TIs and TSMs are closely related to 2D QSH insulators~\cite{WHM_2015}.
They can be built up by stacking 2D QSH insulating layers
along a certain crystal orientation while the QSH
effect can be achieved in an exfoliated, monolayer TI and TSM.
A well-studied example is transition metal
dichalcogenides
$1 \mathrm{T}^{\prime}$-MX$_2$ (M = W, Mo and X = S, Se, Te)~\cite{XFQ_2014}.
This is a new way to search for new layered materials
from the catalogue of topological electronic materials
and study the two-dimensional topological nature
to achieve QSH effect.

In this paper, we study
the structure and electronic properties
of bulk and monolayer AuTe$_2$Cl.
We find that the bulk AuTe$_2$Cl
is a TSM
with a band inversion along $\Gamma-$ Y direction.
While the monolayer is a QSH insulator
with a{ $p_y^{-}-p_z^{+}$ }
band inversion
at the $\bar{\Gamma}$ point in the 2D BZ.
This band inversion is driven by the
confinement effect, which
eliminates the coupling between $p_y$ orbitals
along $y$ direction and pulls down
the anti-bonding state $p_y^{-}$ below the
Fermi level and the bonding state $p_z^{+}$.
Its topological nontrivial nature is demonstrated
through the Wannier center evolution
and surface density of state (DOS) calculations.
Therefore, we present
a new way of exploring
2D QSH insulators
from 3D TSMs.
We expect the predicted AuTe$_2$Cl monolayer
and its QSH effect can be realized in future experiments.

\begin{figure*}
\includegraphics[width=13cm]{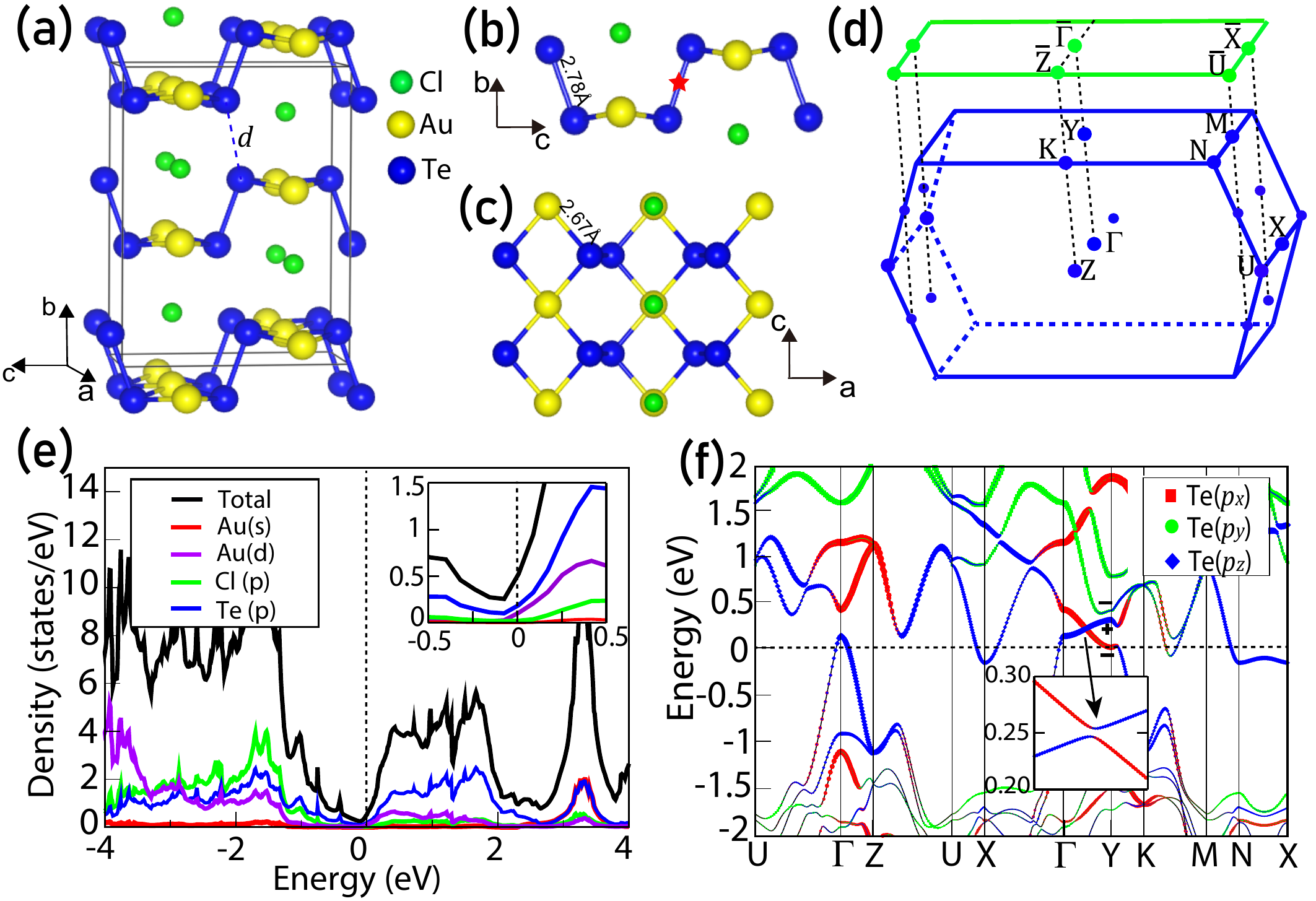}
\caption{
(a) The crystal structure of bulk AuTe$_2$Cl.
It has a layered orthorhombic structure and $d$ is the distance of two adjacent layers.
(b), (c) are the side view and top view of the AuTe$_2$Cl monolayer, respectively.
The red star in (b) denotes an inversion center.
(d) The 3D Brillouin zone
for the bulk and the 2D BZ for the monolayer with high symmetry
$k$ points labeled.
(e) The total DOS and projected DOS
for bulk AuTe$_2$Cl.
(f) The projected band structure for the three $p$ orbitals
of Te in the bulk AuTe$_2$Cl.
The size of the red squares, green circles
and blue diamonds represent
the weight of $p_x$, $p_y$ and $p_z$ orbitals, respectively.
}
\label{fig1}
\end{figure*}

\section{Methods}
First-principles calculations based on density functional
theory (DFT) are carried out by the Vienna \emph{ab initio} simulation
package (VASP)~\cite{PhysRevB.54.11169,PhysRevB.48.13115,PhysRevB.59.1758}.
The exchange-correlation functional
is treated within the local-density approximation (LDA)~\cite{PhysRevB.23.5048}.
A self-consistent field method (tolerance 10$^{-5}$
eV/atom) is employed in conjunction with plane-wave basis sets of
cutoff energy of 500 eV.
Atomic structure optimization is implemented until the remanent Hellmann-Feynman
forces on the ions are less than 0.01 eV/{\AA}.
We use $11\times11\times7$
and $11\times1\times7$ $\Gamma$-centered {\bf k}-meshes to sample the BZ
of bulk and monolayer systems, respectively.
The vacuum layer is set to 15~\AA~to minimize artificial
interaction between layers.
SOC is considered for all calculations.
For monolayer, we also perform hybrid functions calculations
to correct its band structure~\cite{JH_2003}.
The maximum localized Wannier functions
are constructed by using the Wannier90 package~\cite{AAM_2014}.
The topological properties are calculated
by using Wanniertools~\cite{QSW_2018}.

\section{Results and discussion}

The bulk compound AuTe$_2$Cl has been synthesized in experiment~\cite{HMH_1974,BLZ_1981,ZW_2019}, which adopts a layered orthorhombic structure ($D_{2h}$ point group) and lattice constant a = 4.02~{\AA}, b = 11.87~{\AA}, c = 8.77~{\AA} with the nonsymmorphic space group \emph{Cmcm} (No.~63), as shown in
Fig.~1(a). For each layer, a pair of Te atoms is joined to neighboring Au atoms and forms a
Au-Te-Te-Au corrugated net, which is sandwiched by the upper and lower Cl atoms.
Therefore, each Au atom is coordinated with four Te atoms and, likewisely,
each Te atom is coordinated with four Au atoms.
The bond length of Au-Te and Te-Te covalent bond are 2.67~\AA~and
2.78~\AA, respectively.
The short bond length indicates that strong covalent bonds are formed
and, are responsible for the in-plane crystal stability.
The interaction along ${\bf b}$ axis
are mainly contributed
by the Te-Te bond (3.2~\AA) between two adjacent layers [see $d$ in Fig. 1(a)].
The Te-Te bonds along ${\bf b}$ axis
are much weaker than the covalent bonds in the $xz$ plane,
which means
it is experimentally possible to exfoliated monolayer AuTe$_2$Cl
from bulk samples [see Figs.~1(b) and~1(c)].
The interlayer coupling strength of AuTe$_2$Cl
is similar to the cases of GeSe and SnSe~\cite{MT_1990},
but stronger than typical van der Waals materials, such as
graphite~\cite{KBT_1981} and 2H-MoS$_2$~\cite{AS_2010}.

From the projected density of states (PDOS) of bulk AuTe$_2$Cl
shown in Fig.~1(e),
we observe that
the most important orbitals within the energy window
from $-$1.2 to 2.0 eV are the Te $p$ orbitals.
The Au $d$ orbitals and Cl $p$ orbitals mainly
contribute to the states below $-$1.2 eV,
and the Au $s$ orbitals are responsible for the peak at 3.4 eV.
From these observation,
we can understand the electronic transfer and chemical bonding
as follows.
Each Au atom donates a 5$s$ electron to the Cl 3$p$ empty orbitals,
which lowers the total energy and keeps the PDOS of these orbitals away from the Fermi level.
The band structure of bulk system in Fig.~1(f)
reveals a semimetal character,
where the valence band is partially empty resulting in a hole pocket along $\Gamma-$Y,
while the conduction band is partially filled resulting in a electron pocket along X$-$M.
This has been confirmed in transport experiments~\cite{ZW_2019}.
When the SOC is neglected,
there is a Dirac point along $\Gamma-$Y direction
at 0.25~eV above the Fermi level formed
by the $p_x$ and $p_z$ band crossing.
When SOC is considered,
a topological nontrivial gap about 7 meV is opened up.
The parities of $(p_x,p_y,p_z)$ orbitals at $\Gamma$ (and Y) point
are $(-,-,+)$, respectively, as shown in Fig.~1(f).
For the covalent bonding between $p$ orbitals, the bonding
state has the positive parity and the anti-bonding state has the negative parity. Therefore, the $p_{z}$
orbitals are bonding states and the $p_{x}$, $p_{y}$ orbitals are anti-bonding states near the Fermi level.
We notice that
the $p_y$ orbitals have a
stronger dispersion along $\Gamma-$Y direction compared with
$p_x$ and $p_z$ orbitals, which is mainly due to the relatively strong hopping
between $p_{y}$ orbitals along $y$-direction.

Now we focus on the
topological properties of monolayer system.
Due to relatively strong coupling of $p_y$ orbitals
in the $y$ direction,
when bulk AuTe$_2$Cl is exfoliated to
the monolayer,
the confinement effect should be considered.
We simulate the transform
from bulk to monolayer by
increasing the adjacent layer distance $d$.
Fig.~2(a) exhibits the evolution of
conduction band minimum (CBM)
of $p_x$ and $p_y$ orbitals,
and valance band maximum (VBM) of $p_z$ orbitals
at Y point with increasing $d$.
It demonstrates that the CBM of
the $p_y$ orbitals rapidly falls below the $p_x$ orbitals and the Fermi
level. While the CBM of $p_x$ and the VBM of $p_z$ orbitals rise slightly.
When $d > 4.2$~\AA, the energy levels
of three $p$ orbitals reach stationary values, indicating
vanishing of the interlayer coupling along $y$ direction.
Moreover, when the $d$ is large enough, the
band dispersion along $\Gamma-$Y disappears, which means the system
reach its monolayer limit. Hence the topological nature
is determined by the $\bar{\Gamma}$ point in the 2D BZ.

To illustrate the band inversion process in the monolayer AuTe$_2$Cl explicitly,
we start from the Te $p$ orbitals and consider
the effect of chemical bonding and confinement effect for bulk and monolayer AuTe$_2$Cl.
This is schematically depicted in the three stages in Fig.~2(b). Stage I represents
the chemical bonding process between Te atoms.
As analysed above, the states around the Fermi energy are mainly
contributed by bonding states of Te $p_{z}$ orbitals with positive parity
$\xi=+1$ and anti-bonding states of $p_{x}$ and $p_{y}$
orbitals with negative parity $\xi=-1$.
At this stage, the bonding states have lower energy below Fermi level and the anti-bonding
states have higher energy above the Fermi level.
In stage II, the energy levels have changed mainly due to hopping along $y$ direction. The
band inversion occurs between the $p_{x}$ orbitals with parity $\xi=-1$ and the $p_{z}$ orbitals with
parity $\xi=+1$.
Stage III represents the rearrangement of energy levels driven by the confinement effect.
Since $p_y$ orbitals have
larger hopping constants along the $y$ direction, the confinement
effect will influence $p_y$ orbitals much stronger than the $p_x$ and $p_z$ orbitals, and
pulls down the $p_y$ orbitals below the Fermi level.
As a result, the band inversion happens between the bonding state of $p_{z}$ orbitals and the anti-bonding state of $p_{y}$ orbitals in the monolayer AuTe$_2$Cl.

\begin{figure}
\center
\includegraphics[clip,scale=0.410, angle=0]{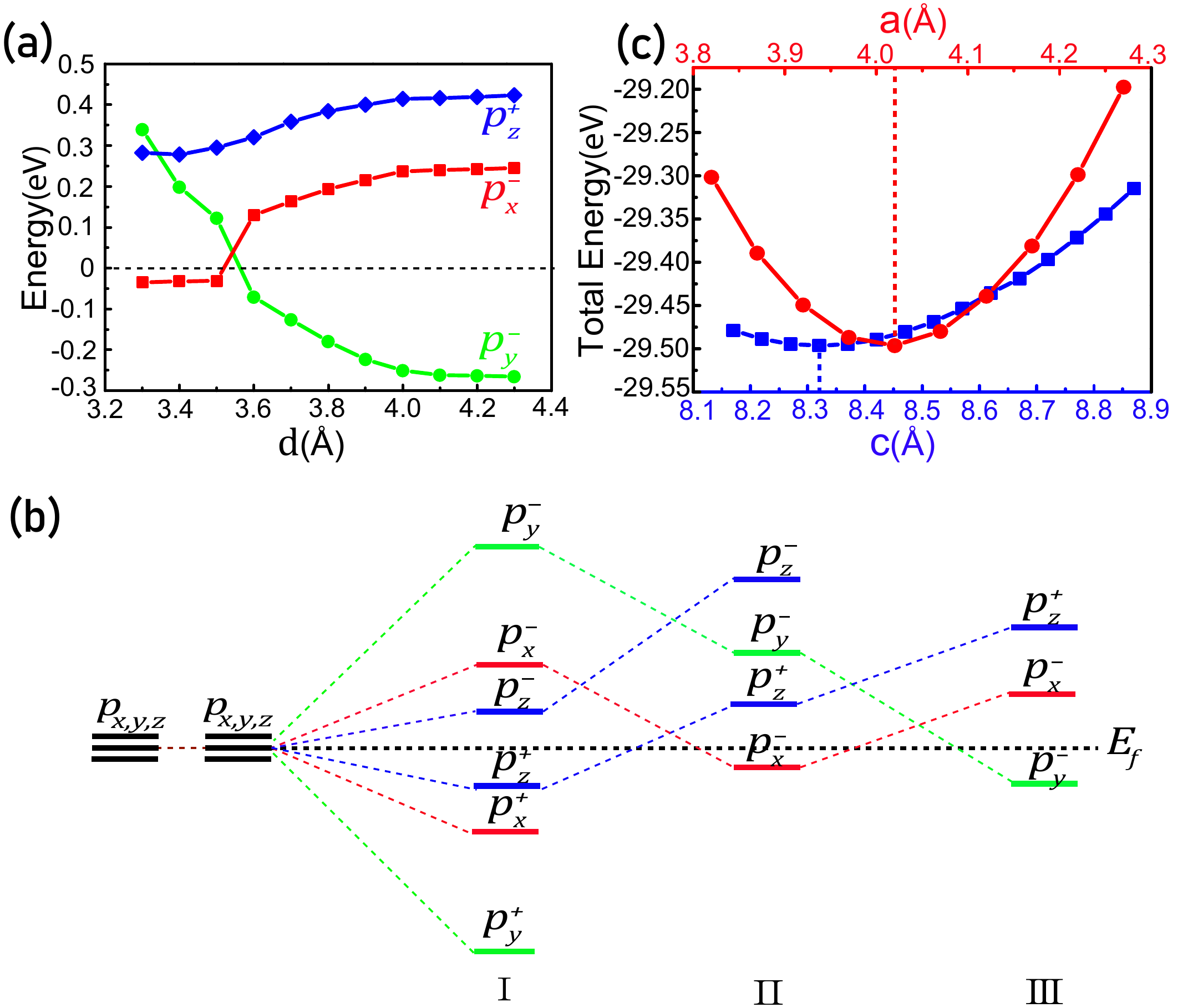}
 \caption{
 (a) The evolution of
the energy level of the three $p$ orbitals of Te, respectively,
with increasing $d$.
(b)
The chemical bonding and confinement effect
on the $p$ orbitals are schematically represented.
I:
The three $p$ orbitals form bonding and anti-bonding states.
II:
The energy level evolution due to hopping along $y$ direction.
The band inversion happens between $p_x^{+}$ and
$p_z^{-}$.
III: The confinement effect lower the $p_y^{-}$ state
and results in a $p_y^{-}-p_{z}^{+}$ band inversion at $\bar{\Gamma}$ point.
(c) The total energy curves with respect to lattice parameters
$a$ and $c$, respectively.
The energy minimum corresponds to $a = 4.02$~\AA~and $c = 8.32$~\AA.
}
\label{fig2}
\end{figure}

\begin{figure*}
\center
\includegraphics[width=13cm]{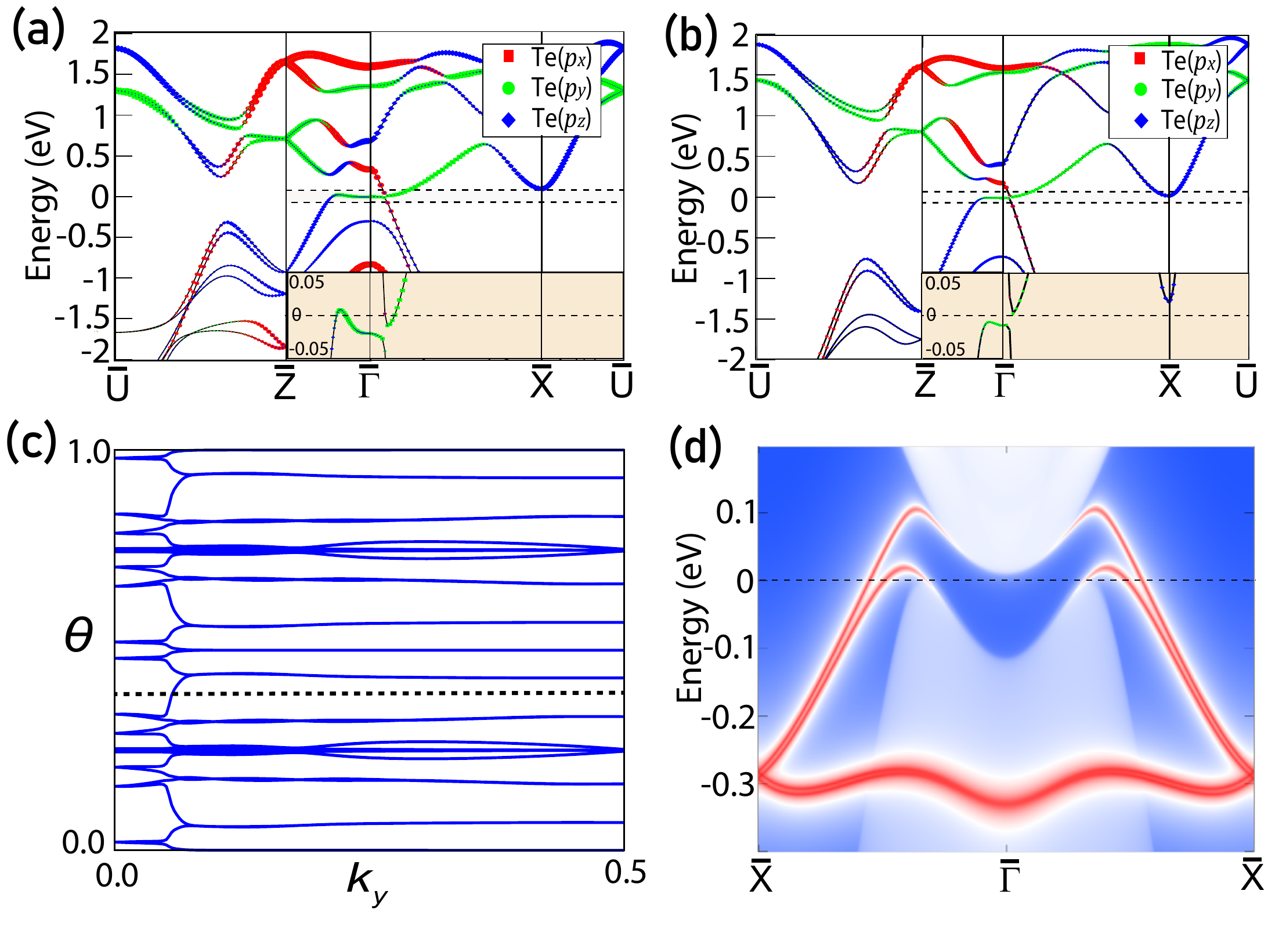}
\caption{
Results for the monolayer AuTe$_2$Cl. (a), (b)
are the band structures
from LDA and hybrid functional calculations, respectively.
In each case, the $p$ orbitals of Te are projected out.
(c) The Wannier center evolution in the $y$-direction
with the SOC calculation.
The reference dash line crosses the evolution
line once which indicates
a topological nontrivial band structure.
(d) The surface DOS along $\bar{\text{X}} - \bar{\Gamma}$
of the 2D BZ.
The red topological edge states are clearly visible.
}
\label{fig3}
\end{figure*}

In order to obtain a stable monolayer structure,
the lattice parameters and ions positions are fully optimized in the LDA calculation.
The total energy with respect to the lattice constants are given in Fig.~2(c).
The lowest total energy corresponds
to $a$ = 4.02~{\AA} and $c$ = 8.32~{\AA}.
The LDA band structure
of monolayer AuTe$_2$Cl is shown in Fig.~3(a).
In the absence of SOC,
the monolayer AuTe$_2$Cl shows semimetal
feature with a Dirac ponit located
on the $\bar{\Gamma}-$$\bar{\text{X}}$ path in the 2D BZ, which confirms the band inversion at $\bar{\Gamma}$ point.
When the SOC is considered,
the Dirac point is gapped out.
Our LDA results suggest
that the monolayer AuTe$_2$Cl is very close
to a insulating state.
The overlap between the CBM and VBM
is about 27~meV.
We further perform the
hybrid functional calculations
using the HSE06 functional.
The HSE06 band structure exhibits a positive band gap about
10~meV at the Fermi level, as shown in Fig.~3(b).
This means when interlayer distance $d$ is increased,
the system undergoes a
topological semimetal to topological insulator phase transition.

Since monolayer AuTe$_2$Cl has
inversion symmetry,
the $Z_2$ topological invariant
can be determined by
the parities of all occupied bands at
the four time-reversal-invariant-momentum (TRIM) points.
According to
Fu-Kane formula,
$(-1)^v = \prod_i \delta_i$ with
$\delta_i = \prod_{m=1}^N \xi_{2m}(\Gamma_i)$.
Here $v$ is the $Z_2$ number and $\xi_{2m}(\Gamma_i)$ is the parity
of the $2m$th occupied band at TRIM ponit $\Gamma_i$~\cite{PhysRevB.76.045302}.
For monolayer AuTe$_2$Cl, the calculated $\delta_i$
for four TRIM points $\bar{\Gamma}$,
$\bar{\text{Y}}$, $\bar{\text{X}}$
and $\bar{\text{U}}$ are given by $-,-,-,+$, respectively.
Hence the $Z_2$ number $v = 1$,
which demonstrates that it is a topological nontrivial insulator.
The topological nontrivial nature
can also be confirmed by the calculated Wannier center evolution
as shown in Fig.~3(c).
The reference dash line crosses the evolution line an odd
number of times in the $y$-direction.
We also calculate the edge spectrum in Fig.~3(d),
in which pair of gapless edge states connects the conduction bands and the valence
bands.
From all these compelling evidences, we thus conclude that
the monolayer AuTe$_2$Cl is a QSH insulator
with a topological band gap about 10~meV.
It should be noted that the 3D bulk AuTe$_2$Cl has been experimentally synthesied~\cite{HMH_1974,BLZ_1981,ZW_2019}.
We expect the monolayer system can be exfoliated from its 3D counterpart,
or from molecular beam epitaxy growth method.

In summary,
we predict that
the monolayer AuTe$_2$Cl
is a QSH insulator with a
topological band gap about 10~meV.
The nontrivial band topology stems from
the band inversion between $p_y$ anti-bonding state and $p_z$
bonding state of Te atoms at $\bar{\Gamma}$ point.
This band inversion is driven by the confinement effect
when 3D bulk system is exfoliated to monolayer thickness.
Since bulk AuTe$_2$Cl has been experimentally synthesized,
we expect the experimental realization of monolayer AuTe$_2$Cl and the predicted QSH effect
are also very promising.
Our work demonstrates
how 2D QSH insulators can be designed from 3D layered TSMs or TIs.
This may further promote the exploration of 2D TIs.

\section{ Acknowlegements }

The authors thank Shuang Jia
for fruitful discussion.
This work
is supported by
the Ministry of Science and Technology
of China (No.~2018YFA0307000) and the National Natural
Science Foundation of China (No.~11874022).

\end{document}